\documentclass[apjl]{emulateapj-rtx4}

\shorttitle{Ba STARS AND OTHER BINARIES IN GCs}
\shortauthors{D'ORAZI ET AL.}

\begin{document}


\title{Ba STARS AND OTHER BINARIES IN FIRST AND SECOND GENERATION STARS IN GLOBULAR CLUSTERS\altaffilmark{1}}


\author{Valentina D'Orazi\altaffilmark{2}}
\author{Raffaele Gratton\altaffilmark{2}}
\author{Sara Lucatello\altaffilmark{2,3}}
\author{Eugenio Carretta\altaffilmark{4}} 
\author{Angela Bragaglia\altaffilmark{4}} 
\author{Anna F. Marino\altaffilmark{5}}

\altaffiltext{1}{Based on observations collected at ESO under
programs 072.D-507 and 073.D-0211}
\altaffiltext{2}{INAF--Osservatorio Astronomico di Padova, 
vicolo dell'Osservatorio 5 , I-35122, Padova, Italy}
\altaffiltext{3}{Excellence Cluster Universe, Technische Universit\"{a}t 
M\"{u}nchen, Boltzmann Str. 2, D-85748, Garching, Germany}
\altaffiltext{4}{INAF--Osservatorio Astronomico di Bologna, via Ranzani 1, 
I-40127, Bologna, Italy}
\altaffiltext{5}{Dipartimento di Astronomia, Universit\`a di Padova, vicolo
dell'Osservatorio 2, I-35122, Padova, Italy}
\email{valentina.dorazi@oapd.inaf.it}


\begin{abstract}
The determination of the Ba abundance in globular cluster (GC) stars is a very powerful test to 
address several issues in the framework of multiple population scenarios. 
We measured the Ba content for a sample of more than 1200 stars in 15
Galactic GCs, using high-resolution FLAMES/Giraffe spectra.
We found no variation in [Ba/Fe] ratios for different stellar populations within each cluster; this means
that low-mass 
asymptotic giant branch stars do not significantly contribute to the intracluster pollution.\\ 
Very interestingly, we obtained that the fraction of Ba stars in first 
generation (FG)  stars is close
to the values derived for field stars ($\sim$2\%); on the other hand, second
generation (SG) stars present a significant lower fraction.
An independent and successful test, based on radial velocity
variations among giant stars in NGC~6121, confirms our finding: the binary fraction
among FG stars is about $\sim$12\% to be compared with  $\sim$1\% of SG stars.
This is an evidence that SG stars formed in a denser environment, where
infant mortality of binary systems was particularly efficient.
\end{abstract}


\section{Introduction}\label{sec:intro}
About 25 years ago, Renzini \& Buzzoni (1986) indicated globular clusters
(GCs) as the best examples in nature of a simple stellar population, 
being defined as an assembly of coeval, initially chemically homogeneous, single
stars. 
However, recent photometric and spectroscopic studies
significantly challenged this traditional paradigm, changing our 
perspective. In a recent survey, focusing on 19 galactic GCs, Carretta et al. (2009b,
hereafter C09)
have convincingly shown that each GC is composed of multiple stellar generations, with the coexistence of both
first (FG, with low Na/Al/N and high O/Mg/C) and second generation (SG) stars
(high Na/Al/N and low O/Mg/C). Briefly, according to Na abundance
and [O/Na] ratios, GC members were classified as FG 
(Primordial (P) stars) and SG
stars, the last one being further divided into Intermediate (I) and Extreme (E; see Carretta et al. 2009b for a detailed
description on PIE classification).
The main results found by Carretta and coworkers can be summarized as follows:
(1) the P population is present in all GCs (at the $\sim$30\% level),
(2) the I population constitutes the bulk (50\%-70\%) of GC stars, and (3)
 the E component is instead not present in all GCs.

 The chemical composition of SG stars indicates that they are formed from FG stars within a limited mass range; hence,
 in order to reproduce the currently observed ratio between FG and SG 
 stars (about 1/3 FG versus 2/3 SG), the original cluster population should have been much larger
 than the current value. In a companion paper, Carretta et al. (2010) indeed concluded that
 during the early epochs of dynamical evolution, a proto-GC should have lost up
 to $\sim$90\% (or even more, see, e.g., Gratton \& Carretta 2010) of 
 its P population. As a consequence they suggested 
 that the P stars in GCs might be the major building blocks of the
 stellar halo.

 In general, the basic idea expressed in Carretta et al. (2010) is that a
 GC, a few 10$^7$yr old, should appear as a rather compact aggregate of 
 stars immersed in
 a loose, wide association and in an even larger expanding gas cloud
 (note that such objects are indeed observed in extragalactic systems, and
 especially in the interacting ones, see, e.g., Vinko et al. 2009).

 For the subsequent evolution, following the approach by
 D'Ercole et al. (2008), Carretta et al. suggested that SG
 stars formed in the central cluster regions, where ejecta of the FG component are channeled into a cooling flow. 
 A small fraction of
 P stars are trapped in the dense, central
 cluster regions formed by SG stars: this is the typical GC that we can observe nowadays.
 
 This proposed scenario is still qualitative, given the lack of more stringent 
 observational evidences to quantitatively constrain such indications.
 Along with uncertainties on cluster formation and early evolution, the nature of polluter stars 
 responsible for the peculiar chemical
 pattern observed in GCs is also still unclear. In fact, 
the presence of Na$-$O (and also Mg$-$Al) anticorrelations also among the main-sequence TO stars 
(Gratton et al. 2001, Cohen et al. 2002, Carretta et al. 2004, 
Pasquini et al. 2005, Lind et al. 2009, D'Orazi et al. 2010) definitively demonstrated that a previous
generation of stars, in which the complete CNO cycle has operated,
originates chemical (anti)correlations found in GCs. 
The nature of the polluters is still debated, the main candidates being intermediate-mass 
asymptotic giant branch (IM$-$AGB; 5$-$8M$_\odot$) stars in hot bottom burning phase (Ventura \& D'Antona 2009), or Fast 
Rotating Massive stars (FRMs; e.g., Decressin et al. 2007), or massive binaries (de Mink et al. 2009).

A complementary approach to proton-capture element
abundance determination, is the study of the $s$-process elements, like e.g., 
barium (see, e.g., Smith 2008). This information can allow to address both (1) the formation and early
evolution mechanisms in GCs and (2) the origin of the massive, previous generation stars
which produced the ejecta for the intracluster enrichment.

The derivation of Ba abundances for a very large
sample of GC stars allows discovering the presence of the so-called Ba stars. 
Spectra of Ba stars are characterized by unusually strong CH, CN, and 
heavy elements (e.g., Ba{\sc ii}, Sr{\sc ii}) features. 
It is now well assessed that Ba stars\footnote{Ba stars can be red giants, but also main sequence or sub-giant stars; 
the fact that to date
all the Ba stars observed
in GCs are giants simply reflects a selection effect.} are long-period, 
single-lined spectroscopic binary systems (McClure 1989):
the nucleosynthetic pattern observed can be attributed 
to now unseen companions, which at earlier epoch were low-mass ($\leq$3M$_{\odot}$) AGB stars 
that transferred processed material
to the surviving visible stars (Luck \& Bond 1991). 
The binary system should originally have had a wide enough separation
 to allow evolution of the primary up to the AGB
(for a discussion on Ba stars in GCs, see Gratton et al. 2004). 

The presence and the fraction of Ba stars within GCs indirectly provide hints 
on binarity properties in dense stellar systems, shedding light on formation and evolution
mechanisms. In this sense, given the quite limited number of known Ba stars in stellar
clusters, large samples are crucial to derive statistically significant information.
 Moreover, the simultaneous
determinations for the very same stars of Na,O and Ba abundances offer the
possibility to unveil possible differences (or analogies) between binary
fraction in FG and SG stars.
Interestingly, determinations of the binary fractions in different stellar generations allow to
infer key information on density conditions of their formation environment.
In fact, it is now well known that the binary incidence is directly related to the environmental 
density (e.g., Lada \& Lada 2003): 
this is a basic constraint for hydrodynamical simulations of GC formation and early evolution.

As far as point (2) is concerned, our work aimed at determining Ba
variation inside each GC as signature of low-mass AGB stars contribution to
pollution scenario. In fact, for stars with masses $\leq$3 M$_\odot$, the third
dredge-up becomes very efficient resulting in an $s$-process element enhancement.

In this Letter we present our results on Ba determination for a sample of 15
GCs, for a grand total of more than 1200 stars. 
The huge database allowed us to obtain very interesting results minimizing the
impact of the observational uncertainties. This is by far the largest survey of Ba
abundances in GC stars available to date. 

\section{Sample and Analysis}\label{sec:analysis}
We analyzed high-quality, high-resolution FLAMES/Giraffe spectra
(Pasquini et al. 2002) for a sample of 1205 red giants 
in 15 Galactic GCs, in order to derive Ba abundances. 
Details on target selections, observations, data reduction, 
along with
$p$-capture elements abundances are given in 
C09.

Using the Kurucz (1993) model atmospheres with the ROSA abundance
code (Gratton 1988), we obtained Ba abundances by measuring equivalent 
widths (EWs) for the only Ba~{\sc ii} feature covered by our spectral setup 
(HR13), namely the line at 6141 \AA. 

Ba lines present hyperfine structure occurring in 
odd isotopes ($^{135}$Ba and $^{137}$Ba); moreover the lines due to the five Ba isotopes 
have small isotopic wavelength shifts. 
However, both these effects are
significant only for Ba~{\sc ii} resonance lines, i.e., 4554 and 4934 \AA, 
while the 6141 \AA~ line is scarcely affected (see, e.g.,  Sneden et al. 1996).
Moreover, the blend with the Fe~{\sc i} line at 
6141.73 \AA~ has a negligible impact at the low metallicity of GCs and 
can be ignored.

Iron abundances were presented in Carretta et al. (2009a), while stellar parameters ($T_{\rm eff}$, log~$g$, $\xi$) were previously derived by
C09, to which we refer also for details on 
error estimates. Here, we just mention that, since we aimed at discovering 
possible star-to-star variations within each GC, systematic errors have a rather small effect. Finally, since this Ba line is near the
saturation regime (with rather high EW values), the dominant source of
(internal) uncertainty
is given by the adopted microturbulences, the final
values of effective temperatures and gravity having only a small impact.

\section{Results and Discussion}\label{sec:disc}
\begin{figure}
\begin{center}
\includegraphics[width=8cm]{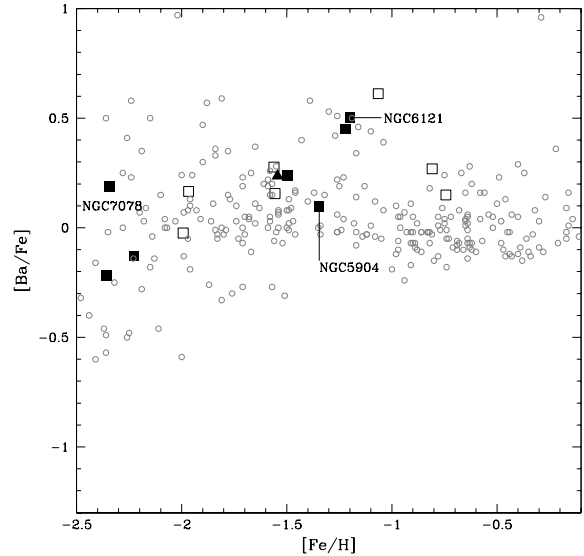}
\caption{Run of average [Ba/Fe] with [Fe/H] for our GCs. Empty and
filled squares are for disk/bulge GC and inner halo GCs; the filled triangle is
for the only outer halo cluster NGC~1904. 
Error bars for [Ba/Fe] and [Fe/H] are similar to the symbol size.
Gray symbols are field stars from the
literature (Fulbright 2000, Burris et al. 2000, Mashonkina \& Gehren 2001, 
Mashonkina et al. 2003). Some GCs are labeled (see the text).}\label{f:bafeh}
\end{center}
\end{figure} 
\begin{figure}
\begin{center}
\includegraphics[width=8cm]{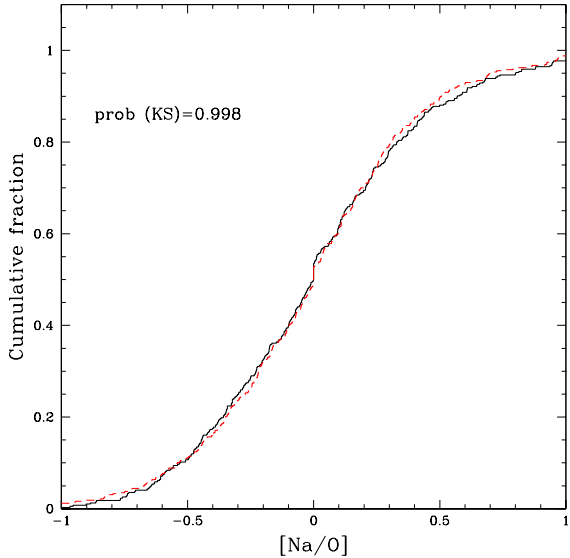}
\caption{[Na/O] distributions for Ba-rich (red) and Ba-poor (black) stars. (A color version of this figure is available
in the online journal.)}\label{f:banao}
\end{center}
\end{figure}
\begin{table}
\caption{Ba Abundances for Our Program Clusters.}\label{t:table1}
\begin{tabular}{lccccr}
\hline\hline
Cluster  &   M$_{\rm V}$ &    [Fe/H] &   [Ba/Fe]    &            rms           &nr$_{\rm stars}$ \\
         &               &      $\pm$err & $\pm$err  &  &  \\
\hline	 
	 	 
         &               &         &         &           &       \\
NGC 104  &   $-$9.42  & $-$0.743$\pm$0.010   &    ~~0.150$\pm$0.014 &	0.150     & 110  \\
NGC 288  &   $-$6.74  & $-$1.219$\pm$0.013   &   ~~0.450$\pm$0.021  &	0.183     & ~75  \\
NGC 1904 &   $-$7.86  & $-$1.544$\pm$0.011   &   ~~0.236$\pm$0.026  &	0.185     & ~49  \\
NGC 3201 &   $-$7.46  & $-$1.495$\pm$0.014   &   ~~0.236$\pm$0.026  &	0.267     & 104  \\
NGC 4590 &   $-$7.35  & $-$2.227$\pm$0.020   &  $-$0.128$\pm$0.033  &	0.316     & ~93  \\
NGC 5904 &   $-$8.81  & $-$1.346$\pm$0.006   &   ~~0.095$\pm$0.021  &	0.210     & 108  \\
NGC 6121 &   $-$7.20  & $-$1.200$\pm$0.007   &   ~~0.502$\pm$0.013  &	0.118     & ~80  \\
NGC 6171 &   $-$7.13  & $-$1.065$\pm$0.020   &   ~~0.612$\pm$0.041  &	0.215     & ~27  \\
NGC 6254 &   $-$7.48  & $-$1.556$\pm$0.014   &   ~~0.156$\pm$0.030  &	0.305     & 105  \\
NGC 6397 &   $-$6.63  & $-$1.993$\pm$0.011   &  $-$0.024$\pm$0.030  &	0.306     & 103  \\
NGC 6752 &   $-$7.73  & $-$1.561$\pm$0.011   &   ~~0.277$\pm$0.017  &	0.176     & 106  \\
NGC 6809 &   $-$7.55  & $-$1.967$\pm$0.012   &   ~~0.166$\pm$0.027  &	0.284     & 112  \\
NGC 6838 &   $-$5.60  & $-$0.808$\pm$0.010   &   ~~0.269$\pm$0.027  &	0.143     & ~29  \\
NGC 7078 &   $-$9.17  & $-$2.341$\pm$0.017   &   ~~0.189$\pm$0.055  &	0.412     & ~56  \\
NGC 7099 &   $-$7.43  & $-$2.359$\pm$0.015   &  $-$0.218$\pm$0.021  &	0.142     & ~48  \\
\hline\hline
\end{tabular}
\end{table}
In Table~\ref{t:table1}, we summarize our results for Ba abundances in our sample GCs; along with metallicity [Fe/H]
as derived by Carretta et al. (2009a), we list the absolute magnitude of the clusters (from Harris 1996), and the mean Ba abundance
for each GC with the corresponding error and the number of stars.
In general, the rms for individual stars from the average [Ba/Fe] ratios, 0.15-0.30 dex, are of the same order of 
magnitude of the expected errors
(about 0.25-0.30 dex),
pointing out that there is no star-to-star variation in Ba content, save for a few exceptional stars (see discussion in the
following section). The only exception is  NGC~7078, which shows 
the largest standard deviation from the mean, i.e., 0.41 dex. For this cluster, 
we derived [Ba/Fe]=0.189$\pm$0.055 to be compared 
with [Ba/Fe]=0.12$\pm$0.04 (rms=0.21) inferred by Sneden et al. (2000) from a sample of 31 giants. 
The quite high [Ba/Fe] ratio along with a significant dispersion for NGC~7078 confirm the previous study  
by Sneden et al. (2000; see also Armosky et al. 1994 for the first suggestion of this trend). However, the [Ba/Eu] ratio derived by 
Sneden et al. (1997, 2000) indicates
 that the heavy element abundances for this GC are compatible with a pure $r$-process pattern: 
 the bimodal distribution for Ba and Eu reveals an $r$-process origin both for Ba and Eu.
For the other GCs for which Ba abundances are available in literature, our estimates are in good agreement with the previous ones. 
Among the others, for the intermediate-metallicity GC NGC~6121 we obtained [Ba/Fe]=0.502$\pm$0.013, 
which is very close to the value derived by Ivans et al. (1999; [Ba/Fe]=0.6 dex). 
Also, we confirm that NGC~6121 shows an overabundance in $s$-process elements with respect to its twin NGC~5904 (for which 
we derived [Ba/Fe]=0.095$\pm$0.021), providing support to previous estimates (see, e.g., Ivans et al. 1999, 2001).
These comparisons with previous determinations prove that no major systematic uncertainties
affect our analysis. 

In Figure~\ref{f:bafeh}, we show the run of [Ba/Fe] as function of metallicity for our 15 GCs, with different symbols 
separating disk/bulge, inner halo, and 
outer halo GCs\footnote{GCs are divided according to a
combination of positional and kinematics criteria. A detailed description of the followed procedure is given in Carretta et al. (2010).}. 
There is not a clear separation between the different Galactic components,
with a large scatter in [Ba/Fe]
ratio at any given metallicity. On the other hand, the well-known raising trend of $s$-process with [Fe/H] is easily recognizable. 
This can be attributed to the AGB stars contribution to the Ba production, which becomes
dominant at metallicity near [Fe/H]$\approx -$1
(see Busso et al. 1999).

None of the clusters present a correlation between Na and Ba abundances (see also James et al. 2004),
indicating that there is no significant contribution from low-mass AGB stars.
We could reach the same conclusion by considering the cumulative distributions 
for Ba-rich and Ba-poor stars:\footnote{For each GC, Ba-rich and Ba-poor stars are defined as stars with [Ba/Fe] larger and smaller,
respectively, than the median value.} we found that  
the two [Na/O] distributions are almost the same, with a Kolmogorov$-$Smirnov test returning a probability of 99.8\% that they are extracted from
the same population (see Figure~\ref{f:banao}).
Also by dividing our sample stars in P, I, E (C09), the three cumulative curves for [Ba/Fe] ratios are very similar:
within $\sim$1.5$\sigma$ the [Ba/Fe] values for FG and SG stars are the same.
\subsection {Ba stars}

\begin{figure}
\begin{center}
\includegraphics[width=8cm]{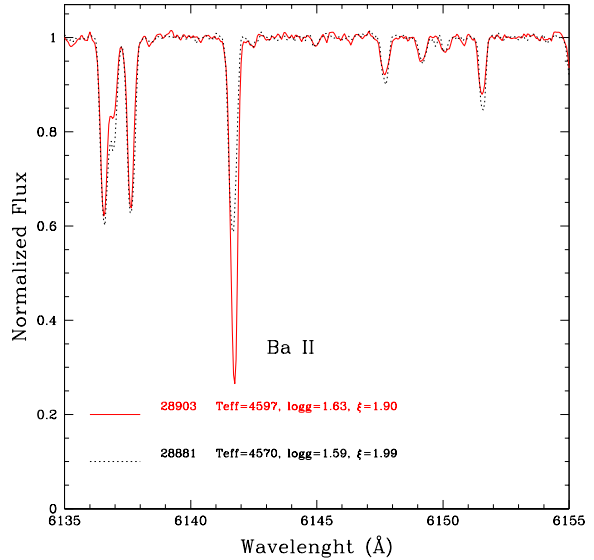}
\caption{Comparison for the spectra of Ba star 28903 (in NGC~6254) with a normal
star with very similar atmospheric parameters (star 28881). Note the strong
differences in Ba~{\sc ii} line strengths.}\label{f:spec}
\end{center}
\end{figure}

Despite the quite large uncertainties ($\approx$0.25 dex), mainly due to the fact that our spectra were not acquired with the 
specific purpose
of performing heavy element abundance analysis, the large statistics lead to the discovery of five
Ba stars. We defined a Ba star if [Ba/Fe] is more than 3.5$\sigma$ above the average of the other stars.
As an example, we compare in Figure~\ref{f:spec} the spectra of two stars (\# 28903 and \# 28881, both in NGC~6254) which have
the same $T_{\rm eff}$, log~$g$ and [Fe/H]; the first is Ba rich, while the latter
has normal Ba value. We found that the EWs of Ba~{\sc ii} lines differ by more than a factor of
two. Given the similarities in atmospheric parameters, different EW strengths must correspond to different Ba content.
For our five Ba stars , other $s$-process elements were determined, whenever possible, as reported in Table~\ref{t:table2}.  
Discarding elements characterized by features
intrinsically weak in our stars, we 
derived abundances for La~{\sc ii} and Nd~{\sc ii}. 
Taking only lines with EWs larger than 10 m\AA, we used the 5769.06 \AA, 
5797.57 \AA, and 
6390.48 \AA~ lines for La~{\sc ii}, while the 5740.86 \AA~ and 5804.00 \AA~ ones for Nd~{\sc ii}.

Over the whole sample of  1205 stars, we derived a fraction of Ba stars of $\sim$0.4\%,
confirming previous suggestions that the fraction of Ba stars in dense stellar systems like GCs is significantly lower than in 
the field ($\sim$2\%, Luck \& Bond
1991).
Furthermore, Gratton et al. (2004) argued that all known CH and Ba stars in 
GCs are in low concentration clusters:   
this is probably because in highly concentrated clusters, the possible 
progenitor systems are destroyed on a timescale shorter than the evolution of intermediate low-mass stars (Giersz \& Spurzem 2000, see also 
Gratton et al. 2004).
However,  we found no relationship between Ba stars and cluster concentration, with Ba stars present in both NGC~288  and
47Tuc (NGC~104), which have central concentration parameters of c=0.96 and c=2.03, respectively (see Harris 1996). 
In addition, we did not find any correlation with
metallicity and absolute magnitude.

In Table~\ref{t:table2}, along with heavy elements, we report for our Ba stars also the abundances for Na and O measured by C09.
Four out of the five Ba stars detected in our sample belong to the 
P population. This means that if we compute the fraction of Ba stars only considering the P stars,
the percentage is $\sim$ 2\%,  the same number observed in field stars. 
This interesting finding supports our scenario for GC formation, where the SG stars were born in a denser environment
(generated by the cooling flow), while the primordial population formed as an extended and loose association.  
The environment conditions could have influenced the wide binaries population, since in high-density conditions
 the high rate of collisions significantly contribute to the destruction of such systems, while in sparse environment binaries 
 can survive in a larger number.

As an independent check, to confirm the suggestion from Ba stars, we also analyzed radial velocity variations for a hundred  giant stars in
NGC~6121, looking for spectroscopic binaries (more details will be given in a separate paper, S. Lucatello et al., 2010, in preparation). We chose stars for which at least two radial velocity determinations were
available, either within our observations program (time span of $\Delta$t=3 months) or combining our data and the observations carried out
by Marino et al. (2008; $\Delta$t=2 years).
We discovered five binary stars on a grand total of 102, which means a fraction of $\sim$5\%: our result is in good
agreement with the binary fraction for this GC found by A. P. Milone et al. (2010, in preparation), i.e., 6.0\%$\pm$0.4\% (by considering only systems 
with a mass ratio $q$=M1/M2 $>$0.5, see Milone et al. 2008 for further details on this point.)

Considering the O and Na content for these binaries, we found that four out of five belong to the P component
(with low Na content). We performed a statistical test and the probability that 
this finding can be obtained by chance is only of $\sim$5\%.
Assuming a fraction of 34\%$\pm$6\% for FG and  66\%$\pm$8\% for SG stars\footnote{The 102 stars of our sample were divided according to the Na abundance, 
with FG having [Na/Fe]$<$0.28 dex (see the Introduction). 
Our estimate is in good agreement with the values of 30\%$\pm$5\% and 70\%$\pm$8\% found by C09,
respectively for FG and SG stars. The main difference between our computation and the one by C09 is that 
we divided FG and SG using only 
Na abundances, while C09 considered both Na and O contents.}, we
hence derived a value of $\sim$12\% for the frequency of binaries among
P stars, and only $\sim$1\% for SG stars.

Finally, we investigated whether there is a trend between the fraction of FG stars within a GC and the number of Blue
Stragglers (BSS), since BSS seem to have predominantly an origin as P binaries, with only a
small percentage formed through different mechanisms, e.g., collisions (see, e.g., Moretti et al. 2008). 
By comparing the fraction of P stars in 14 GCs in common between our previous paper (C09) and the one by Moretti et al., 
we indeed find that there is a positive correlation between the number of FG stars and the BSS fraction (the linear correlation 
coefficient has a
significance level of 97.5\%). 
This is a further confirmation of what derived from both Ba stars and 
spectroscopic binaries in NGC~6121.

Our result is yet another observational 
evidence, along with metallicity distribution, mass and location, of the similarities between the primordial component 
of GCs and the field stars. 
In a forthcoming paper (R. G. Gratton et al., 2010, in preparation), we will widely discuss this issue. 
Since second generation stars are nowadays dominant in GCs, our result could explain why 
binary fractions in GCs are smaller compared to the field (e.g.,
Pryor et al. 1989; Hut et al. 1992; Cote et al. 1996).
Note that at variance with the field, different processes determine the relative frequency of binary systems in stellar
clusters; in GCs binaries are continuously formed and 
destroyed during the cluster evolution as the  result of collisional interactions between binaries and single stars
(see, e.g., Sollima et al. 2009). However, the cluster regions sampled by FLAMES are generally quite far from the center, where
dynamical interactions are much more frequent. This might explain why we can observe these differences between stars 
belonging to different stellar generations some 10-13 Gyr after GC formation.

\begin{table*}
\begin{center}
\caption{Abundances for Ba-rich stars.}
\label{t:table2}
\begin{tabular}{lcccccc}
\hline\hline
Star & cluster &  [O/Fe] & [Na/Fe] &  [Ba$_{\rm \sc II}$/Fe] & [La$_{\rm \sc II}$/Fe] & [Nd$_{\rm \sc II}$/Fe]\\
\hline
         &          &                    &                 &                       &                 &     \\
~~30952  & NGC ~104 & ~~~0.336$\pm$0.046 & 0.343$\pm$0.023 &  0.508$\pm$0.150 & ~~---  & ~~---\\
 200095  & NGC ~288 & $-$0.045$\pm$0.088 & 0.079$\pm$0.053 &  1.115$\pm$0.210 & 1.209$\pm$0.044  & 1.035$\pm$0.030\\
~~28903	 & NGC 6254 & ~~~0.428$\pm$0.036 & 0.203$\pm$0.031 &  1.464$\pm$0.160 & 0.281$\pm$0.150   & ~~--- \\
 608024	 & NGC 6397 & ~~~0.387$\pm$0.048 & ~~---           &  1.040$\pm$0.344 & ~~---  & ~~---\\
~~38291	 & NGC 6752 & ~~~0.253$\pm$0.043 & 0.126$\pm$0.038 &  1.261$\pm$0.182 & ~~---  & ~~---\\
\hline\hline
\end{tabular}
\end{center}
\end{table*}

\section{Summary and Future perspectives}\label{sec:sum}
In this Letter, we report our results on Ba abundance determination for a sample
of more than 1200 giants in 15 Galactic GCs. We provide the largest,
homogeneous analysis of this kind available in literature to date.

We found that there is no star-to-star variation in Ba content within each cluster, with the
standard deviation from the mean compatible with the expected uncertainties 
(the only exception being the low-metallicity NGC~7078, for which, however, a pure $r$-process pattern has 
been proposed by several authors). 
Our results suggest that there is no detectable contribution to intra-cluster
pollution from low-mass AGB stars, whose products are $s$-process elements enhanced due to
the efficiency of the third dredge-up.

We discovered the presence of five Ba stars, confirming that the fraction of 
these originally quite wide binary systems is lower in GCs than in the field. 
We also found that four out of the five Ba stars belong to
the P population: this likely reflects the different environmental conditions
where the two stellar generations formed. This new result supports
the formation/evolution scenario for GC in which the SG stars were born at the
cluster center, where the infant mortality of long-period binaries is
significantly higher.

To have a further validation of this piece of evidence, we also searched for
binary stars in NGC~6121: for this cluster, we determined rather large 
radial velocity variations for five stars on a total of $\sim$100. Also in this
case we obtained that only one binary  belongs to SG stars, and that the fraction of binary stars among the P
component reaches a value of $\sim$12\%, to be compared with the small percentage in SG, i.e., $\sim$1\%.

In the future, we will focus on the determination of radial velocity variations among a large sample of GCs and of
stars within each GC. We aim at inferring binarity properties in the FG and SG of GCs, as
excellent tracer of the environmental conditions where GCs formed and evolved.

\acknowledgments
This work has been funded by PRIN INAF 2007 "{\it Multiple Stellar 
Populations in Globular Clusters"} 
and DFG cluster of excellence "Origin and Structure of the
Universe".  A.P. Milone is kindly acknowledged for having provided photometric information in advance of publication.
We thank the anonymous referee for her/his very positive comments and helpful suggestions.

{}

\end{document}